

%
%


\def\famname{
 \textfont0=\textrm \scriptfont0=\scriptrm
 \scriptscriptfont0=\sscriptrm
 \textfont1=\textmi \scriptfont1=\scriptmi
 \scriptscriptfont1=\sscriptmi
 \textfont2=\textsy \scriptfont2=\scriptsy \scriptscriptfont2=\sscriptsy
 \textfont3=\textex \scriptfont3=\textex \scriptscriptfont3=\textex
 \textfont4=\textbf \scriptfont4=\scriptbf \scriptscriptfont4=\sscriptbf
 \skewchar\textmi='177 \skewchar\scriptmi='177
 \skewchar\sscriptmi='177
 \skewchar\textsy='60 \skewchar\scriptsy='60
 \skewchar\sscriptsy='60
 \def\rm{\fam0 \textrm} \def\bf{\fam4 \textbf}}
\def\sca#1{scaled\magstep#1} \def\scah{scaled\magstephalf} 
\def\twelvepoint{
 \font\textrm=cmr12 \font\scriptrm=cmr8 \font\sscriptrm=cmr6
 \font\textmi=cmmi12 \font\scriptmi=cmmi8 \font\sscriptmi=cmmi6 
 \font\textsy=cmsy10 \sca1 \font\scriptsy=cmsy8
 \font\sscriptsy=cmsy6
 \font\textex=cmex10 \sca1
 \font\textbf=cmbx12 \font\scriptbf=cmbx8 \font\sscriptbf=cmbx6
 \font\it=cmti12
 \font\sectfont=cmbx12 \sca1
 \font\sectmath=cmmib10 \sca2
 \font\sectsymb=cmbsy10 \sca2
 \font\refrm=cmr10 \scah \font\refit=cmti10 \scah
 \font\refbf=cmbx10 \scah
 \def\twelverm{\textrm} \def\twelveit{\it} \def\twelvebf{\textbf}
 \famname \textrm 
 \advance\voffset by .06in \advance\hoffset by .28in
 \normalbaselineskip=17.5pt plus 1pt \baselineskip=\normalbaselineskip
 \parindent=21pt
 \setbox\strutbox=\hbox{\vrule height10.5pt depth4pt width0pt}}


\catcode`@=11

{\catcode`\'=\active \def'{{}^\bgroup\prim@s}}

\def\screwcount{\alloc@0\count\countdef\insc@unt}   
\def\screwdimen{\alloc@1\dimen\dimendef\insc@unt} 
\def\screwbox{\alloc@4\box\chardef\insc@unt}

\catcode`@=12


\overfullrule=0pt			
\vsize=9in \hsize=6in
\lineskip=0pt				
\abovedisplayskip=1.2em plus.3em minus.9em 
\belowdisplayskip=1.2em plus.3em minus.9em	
\abovedisplayshortskip=0em plus.3em	
\belowdisplayshortskip=.7em plus.3em minus.4em	
\parindent=21pt
\setbox\strutbox=\hbox{\vrule height10.5pt depth4pt width0pt}
\def\makefootline{\baselineskip=30pt \line{\the\footline}}
\footline={\ifnum\count0=1 \hfil \else\hss\twelverm\folio\hss \fi}
\pageno=1


\def\put(#1,#2)#3{\screwdimen\unit  \unit=1in
	\vbox to0pt{\kern-#2\unit\hbox{\kern#1\unit
	\vbox{#3}}\vss}\nointerlineskip}


\def\\{\hfil\break}
\def\newpage{\vfill\eject}
\def\center{\leftskip=0pt plus 1fill \rightskip=\leftskip \parindent=0pt
 \def\textindent##1{\par\hangindent21pt\footrm\noindent\hskip21pt
 \llap{##1\enspace}\ignorespaces}\par}
\def\unnarrower{\leftskip=0pt \rightskip=\leftskip}


\def\vol#1 {{\refbf#1} }		 


\def\NP #1 {{\refit Nucl. Phys.} {\refbf B{#1}} }
\def\PL #1 {{\refit Phys. Lett.} {\refbf{#1}} }
\def\PR #1 {{\refit Phys. Rev. Lett.} {\refbf{#1}} }
\def\PRD #1 {{\refit Phys. Rev.} {\refbf D{#1}} }


\hyphenation{pre-print}
\hyphenation{quan-ti-za-tion}

%
%


\def\oonoo#1#2#3{\vbox{\ialign{##\crcr
	\hfil\hfil\hfil{$#3{#1}$}\hfil\crcr\noalign{\kern1pt\nointerlineskip}
	$#3{#2}$\crcr}}}
\def\oon#1#2{\mathchoice{\oonoo{#1}{#2}{\displaystyle}}
	{\oonoo{#1}{#2}{\textstyle}}{\oonoo{#1}{#2}{\scriptstyle}}
	{\oonoo{#1}{#2}{\scriptscriptstyle}}}
\def\dt#1{\oon{\hbox{\bf .}}{#1}}  
\def\ddt#1{\oon{\hbox{\bf .\kern-1pt.}}#1}    
\def\slap#1#2{\setbox0=\hbox{$#1{#2}$}
	#2\kern-\wd0{\hfuzz=1pt\hbox to\wd0{\hfil$#1{/}$\hfil}}}
\def\sla#1{\mathpalette\slap{#1}}                
\def\bop#1{\setbox0=\hbox{$#1M$}\mkern1.5mu
	\lower.02\ht0\vbox{\hrule height0pt depth.06\ht0
	\hbox{\vrule width.06\ht0 height.9\ht0 \kern.9\ht0
	\vrule width.06\ht0}\hrule height.06\ht0}\mkern1.5mu}
\def\bo{{\mathpalette\bop{}}}                        
\def~{\widetilde} 
\mathcode`\*="702A                  
\def\in{\relax\ifmmode\mathchar"3232\else{\refit in\/}\fi} 
\def\f#1#2{{\textstyle{#1\over#2}}}	   
\def\half{{\textstyle{1\over{\raise.1ex\hbox{$\scriptstyle{2}$}}}}}

\def\Gamma{\mathchar"0100}
\def\Delta{\mathchar"0101}
\def\Theta{\mathchar"0102}
\def\Lambda{\mathchar"0103}
\def\Xi{\mathchar"0104}
\def\Pi{\mathchar"0105}
\def\Sigma{\mathchar"0106}
\def\Upsilon{\mathchar"0107}
\def\Phi{\mathchar"0108}
\def\Psi{\mathchar"0109}
\def\Omega{\mathchar"010A}

\catcode128=13 \def €{\"A}                 
\catcode129=13 \def {\AA}                 
\catcode130=13 \def '{\c}           	   
\catcode131=13 \def ƒ{\'E}                   
\catcode132=13 \def "{\~N}                   
\catcode133=13 \def …{\"O}                 
\catcode134=13 \def †{\"U}                  
\catcode135=13 \def ‡{\'a}                  
\catcode136=13 \def ˆ{\`a}                   
\catcode137=13 \def ‰{\^a}                 
\catcode138=13 \def Š{\"a}                 
\catcode139=13 \def ‹{\~a}                   
\catcode140=13 \def Œ{\alpha}            
\catcode141=13 \def {\chi}                
\catcode142=13 \def Ž{\'e}                   
\catcode143=13 \def {\`e}                    
\catcode144=13 \def {\^e}                  
\catcode145=13 \def '{\"e}                
\catcode146=13 \def '{\'\i}                 
\catcode147=13 \def "{\`\i}                  
\catcode148=13 \def "{\^\i}                
\catcode149=13 \def •{\"\i}                
\catcode150=13 \def –{\~n}                  
\catcode151=13 \def —{\'o}                 
\catcode152=13 \def ˜{\`o}                  
\catcode153=13 \def ™{\^o}                
\catcode154=13 \def š{\"o}                 
\catcode155=13 \def ›{\~o}                  
\catcode156=13 \def œ{\'u}                  
\catcode157=13 \def {\`u}                  
\catcode158=13 \def ž{\^u}                
\catcode159=13 \def Ÿ{\"u}                
\catcode160=13 \def  {\tau}               
\catcode161=13 \mathchardef ¡="2203     
\catcode162=13 \def ¢{\oplus}           
\catcode163=13 \def £{\relax\ifmmode\to\else\itemize\fi} 
\catcode164=13 \def ¤{\subset}	  
\catcode165=13 \def ¥{\infty}           
\catcode166=13 \def ¦{\mp}                
\catcode167=13 \def §{\sigma}           
\catcode168=13 \def ¨{\rho}               
\catcode169=13 \def ©{\gamma}         
\catcode170=13 \def ª{\leftrightarrow} 
\catcode171=13 \def «{\relax\ifmmode\acute\else\expandafter\'\fi}
\catcode172=13 \def ¬{\relax\ifmmode\expandafter\ddt\else\expandafter\"\fi}
\catcode173=13 \def ­{\equiv}            
\catcode174=13 \def ®{\approx}          
\catcode175=13 \def ¯{\Omega}          
\catcode176=13 \def °{\otimes}          
\catcode177=13 \def ±{\ne}                 
\catcode178=13 \def ²{\le}                   
\catcode179=13 \def ³{\ge}                  
\catcode180=13 \def ´{\upsilon}          
\catcode181=13 \def µ{\mu}                
\catcode182=13 \def ¶{\delta}             
\catcode183=13 \def ·{\epsilon}          
\catcode184=13 \def ¸{\Pi}                  
\catcode185=13 \def ¹{\pi}                  
\catcode186=13 \def º{\beta}               
\catcode187=13 \def »{\partial}           
\catcode188=13 \def ¼{\nobreak\ }       
\catcode189=13 \def ½{\zeta}               
\catcode190=13 \def ¾{\sim}                 
\catcode191=13 \def ¿{\omega}           
\catcode192=13 \def À{\dt}                     
\catcode193=13 \def Á{\gets}                
\catcode194=13 \def Â{\lambda}           
\catcode195=13 \def Ã{\nu}                   
\catcode196=13 \def Ä{\phi}                  
\catcode197=13 \def Å{\xi}                     
\catcode198=13 \def Æ{\psi}                  
\catcode199=13 \def Ç{\int}                    
\catcode200=13 \def È{\oint}                 
\catcode201=13 \def É{\relax\ifmmode\cdot\else\vol\fi}    
\catcode202=13 \def Ê{\relax\ifmmode\,\else\thinspace\fi}
\catcode203=13 \def Ë{\`A}                      
\catcode204=13 \def Ì{\~A}                      
\catcode205=13 \def Í{\~O}                      
\catcode206=13 \def Î{\Theta}              
\catcode207=13 \def Ï{\theta}               
\catcode208=13 \def Ð{\relax\ifmmode\bar\else\expandafter\=\fi}
\catcode209=13 \def Ñ{\overline}             
\catcode210=13 \def Ò{\langle}               
\catcode211=13 \def Ó{\relax\ifmmode\{\else\ital\fi}      
\catcode212=13 \def Ô{\rangle}               
\catcode213=13 \def Õ{\}}                        
\catcode214=13 \def Ö{\sla}                      
\catcode215=13 \def ×{\relax\ifmmode\check\else\expandafter\v\fi}
\catcode216=13 \def Ø{\"y}                     
\catcode217=13 \def Ù{\"Y}  		    
\catcode218=13 \def Ú{\Leftarrow}       
\catcode219=13 \def Û{\Leftrightarrow}       
\catcode220=13 \def Ü{\relax\ifmmode\Rightarrow\else\sect\fi}
\catcode221=13 \def Ý{\sum}                  
\catcode222=13 \def Þ{\prod}                 
\catcode223=13 \def ß{\widehat}              
\catcode224=13 \def à{\pm}                     
\catcode225=13 \def á{\nabla}                
\catcode226=13 \def â{\quad}                 
\catcode227=13 \def ã{\in}               	
\catcode228=13 \def ä{\star}      	      
\catcode229=13 \def å{\sqrt}                   
\catcode230=13 \def æ{\^E}			
\catcode231=13 \def ç{\Upsilon}              
\catcode232=13 \def è{\"E}    	   	 
\catcode233=13 \def é{\`E}               	  
\catcode234=13 \def ê{\Sigma}                
\catcode235=13 \def ë{\Delta}                 
\catcode236=13 \def ì{\Phi}                     
\catcode237=13 \def í{\`I}        		   
\catcode238=13 \def î{\iota}        	     
\catcode239=13 \def ï{\Psi}                     
\catcode240=13 \def ð{\times}                  
\catcode241=13 \def ñ{\Lambda}             
\catcode242=13 \def ò{\cdots}                
\catcode243=13 \def ó{\^U}			
\catcode244=13 \def ô{\`U}    	              
\catcode245=13 \def õ{\bo}                       
\catcode246=13 \def ö{\relax\ifmmode\hat\else\expandafter\^\fi}
\catcode247=13 \def÷{\relax\ifmmode\tilde\else\expandafter\~\fi}
\catcode248=13 \def ø{\ll}                         
\catcode249=13 \def ù{\gg}                       
\catcode250=13 \def ú{\eta}                      
\catcode251=13 \def û{\kappa}                  
\catcode252=13 \def ü{\half}     		 
\catcode253=13 \def ý{\Gamma} 		
\catcode254=13 \def þ{\Xi}   			
\catcode255=13 \def ÿ{\relax\ifmmode{}^{\dagger}{}\else\dag\fi}


\def\ital#1Õ{{\it#1\/}}	     
\def\un#1{\relax\ifmmode\underline#1\else $\underline{\hbox{#1}}$
	\relax\fi}

\def\roonoo#1#2#3{\vbox{\ialign{##\crcr
	\hfil{$#3{#1}$}\hfil\crcr\noalign{\kern1pt\nointerlineskip}
	$#3{#2}$\crcr}}}

\def\tdt#1{\oon{\hbox{\bf .\kern-1pt.\kern-1pt.}}#1}   
\def\({\eqno(}
\def\li{\openup1\jot \eqalignno}


\def\õ#1{
	\screwcount\num
	\num=1
	\screwdimen\downsy
	\downsy=-1.5ex
	\mkern-3.5mu
	õ
	\loop
	\ifnum\num<#1
	\llap{\raise\num\downsy\hbox{$õ$}}
	\advance\num by1
	\repeat}
\def\upõ#1#2{\screwcount\numup
	\numup=#1
	\advance\numup by-1
	\screwdimen\upsy
	\upsy=.75ex
	\mkern3.5mu
	\raise\numup\upsy\hbox{$#2$}}



\newcount\marknumber	\marknumber=1
\newcount\countdp \newcount\countwd \newcount\countht 

%
%
\ifx\pdfoutput\undefined
\def\rgboo#1{}
\input epsf

\def\postscript#1{\special{" #1}}		
\postscript{
	/bd {bind def} bind def
	/fsd {findfont exch scalefont def} bd
	/sms {setfont moveto show} bd
	/ms {moveto show} bd
	/pdfmark where		
	{pop} {userdict /pdfmark /cleartomark load put} ifelse
	[ /PageMode /UseOutlines		
	/DOCVIEW pdfmark}
\def\bookmark#1#2{\postscript{		
	[ /Dest /MyDest\the\marknumber /View [ /XYZ null null null ] /DEST pdfmark
	[ /Title (#2) /Count #1 /Dest /MyDest\the\marknumber /OUT pdfmark}%
	\advance\marknumber by1}
\def\pdfklink#1#2{%
	\hskip-.25em\setbox0=\hbox{#1}%
		\countdp=\dp0 \countwd=\wd0 \countht=\ht0%
		\divide\countdp by65536 \divide\countwd by65536%
			\divide\countht by65536%
		\advance\countdp by1 \advance\countwd by1%
			\advance\countht by1%
		\def\linkdp{\the\countdp} \def\linkwd{\the\countwd}%
			\def\linkht{\the\countht}%
	\postscript{
		[ /Rect [ -1.5 -\linkdp.0 0\linkwd.0 0\linkht.5 ] 
		/Border [ 0 0 0 ]
		/Action << /Subtype /URI /URI (#2) >>
		/Subtype /Link
		/ANN pdfmark}{\rgb{1 0 0}{#1}}}
%
%
\else
\def\rgboo#1{\pdfliteral{#1 rg #1 RG}}

\pdfcatalog{/PageMode /UseOutlines}		
\def\bookmark#1#2{
	\pdfdest num \marknumber xyz
	\pdfoutline goto num \marknumber count #1 {#2}
	\advance\marknumber by1}
\def\pdfklink#1#2{%
	\noindent\pdfstartlink user
		{/Subtype /Link
		/Border [ 0 0 0 ]
		/A << /S /URI /URI (#2) >>}{\rgb{1 0 0}{#1}}%
	\pdfendlink}
\fi

\def\rgbo#1#2{\rgboo{#1}#2\rgboo{0 0 0}}
\def\rgb#1#2{\mark{#1}\rgbo{#1}{#2}\mark{0 0 0}}
\def\pdflink#1{\pdfklink{#1}{#1}}
\def\xxxlink#1{\pdfklink{[arXiv:#1]}{http://arXiv.org/abs/#1}}

\catcode`@=11

\def\wlog#1{}	


\def\makeheadline{\vbox to\z@{\vskip-36.5\p@
	\line{\vbox to8.5\p@{}\the\headline%
	\ifnum\pageno=\z@\rgboo{0 0 0}\else\rgboo{\topmark}\fi%
	}\vss}\nointerlineskip}
\headline={
	\ifnum\pageno=\z@
		\hfil
	\else
		\ifnum\pageno<\z@
			\ifodd\pageno
				\tenrm\romannumeral-\pageno\hfil\lefthead\hfil
			\else
				\tenrm\hfil\righthead\hfil\romannumeral-\pageno
			\fi
		\else
			\ifodd\pageno
				\tenrm\hfil\righthead\hfil\number\pageno
			\else
				\tenrm\number\pageno\hfil\lefthead\hfil
			\fi
		\fi
	\fi}

\catcode`@=12

\def\righthead{\hfil} \def\lefthead{\hfil}
\nopagenumbers


\def\chrulefill{\rgb{1 0 0}{\hrulefill}}
\def\cdotfill{\rgb{1 0 0}{\dotfill}}
\newcount\area	\area=1
\newcount\cross	\cross=1
\def\volume#1\par{\newpage\noindent{\biggest{\rgb{1 .5 0}{#1}}}
	\par\nobreak\bigskip\medskip\area=0}
\def\chapskip{\par\ifnum\area=0\bigskip\medskip\goodbreak
	\else\newpage\fi}
\def\chapy#1{\area=1\cross=0
	\xdef\lefthead{\rgbo{1 0 .5}{#1}}\vbox{\biggerer\offinterlineskip
	\line{\chrulefill¼\hphantom{\lefthead}\chrulefill}
	\line{\chrulefill¼\lefthead\chrulefill}}\par\nobreak\medskip}
\def\chap#1\par{\chapskip\bookmark3{#1}\chapy{#1}}
\def\sectskip{\par\ifnum\cross=0\bigskip\medskip\goodbreak
	\else\newpage\fi}
\def\secty#1{\cross=1
	\xdef\righthead{\rgbo{1 0 1}{#1}}\vbox{\bigger\offinterlineskip
	\line{\cdotfill¼\hphantom{\righthead}\cdotfill}
	\line{\cdotfill¼\righthead\cdotfill}}\par\nobreak\medskip}
\def\sect#1 #2\par{\sectskip\bookmark{#1}{#2}\secty{#2}}
\def\subsectskip{\par\ifdim\lastskip<\medskipamount
	\bigskip\medskip\goodbreak\else\nobreak\fi}
\def\subsecty#1{\noindent{\sectfont{\rgbo{.5 0 1}{#1}}}\par\nobreak\medskip}
\def\subsect#1\par{\subsectskip\bookmark0{#1}\subsecty{#1}}
\long\def\x#1 #2\par{\hangindent2\parindent%
\mark{0 0 1}\rgboo{0 0 1}{\bf Exercise #1}\\#2%
\par\rgboo{0 0 0}\mark{0 0 0}}
\def\refs{\bigskip\noindent{\bf \rgbo{0 .5 1}{REFERENCES}}\par\nobreak\medskip
	\frenchspacing \parskip=0pt \refrm \baselineskip=1.23em plus 1pt
	\def\ital##1Õ{{\refit##1\/}}}
\long\def\twocolumn#1#2{\hbox to\hsize{\vtop{\hsize=2.9in#1}
	\hfil\vtop{\hsize=2.9in #2}}}


\twelvepoint
\font\bigger=cmbx12 \sca2
\font\biggerer=cmb10 \sca5
\font\biggest=cmssdc10 scaled 3583
 \sca5

 \sca3


\def Ü{\relax\ifmmode\Rightarrow\else\expandafter\subsect\fi}
\def Û{\relax\ifmmode\Leftrightarrow\else\expandafter\sect\fi}
\def Ú{\relax\ifmmode\Leftarrow\else\expandafter\chap\fi}

\def\itemize#1 {\item{\bf#1}}
\def\itemizze#1 {\itemitem{\bf#1}}
\def\itemutem{\par\indent\indent \hangindent3\parindent \textindent}
\def\itemizzze#1 {\itemutem{\bf#1}}
\def ª{\relax\ifmmode\leftrightarrow\else\itemizze\fi}
\def Á{\relax\ifmmode\gets\else\itemizzze\fi}

\def\¢{\ominus}
      \def\D{{\cal D}}

  \def\P{{\cal P}}

\def\Ä{\varphi}  \def\¿{\varpi}	\def\Ï{\vartheta}

\def ò{\relax\ifmmode\cdots\else\dotfill\fi}


\def\cvrule{\rgbo{0 .5 1}{\vrule}}
\def\chrule{\rgbo{0 .5 1}{\hrule}}
\def\boxit#1{\leavevmode\thinspace\hbox{\cvrule\vtop{\vbox{\chrule%
	\vskip3pt\kern1pt\hbox{\vphantom{\bf/}\thinspace\thinspace%
	{\bf#1}\thinspace\thinspace}}\kern1pt\vskip3pt\chrule}\cvrule}%
	\thinspace}
\def\Boxit#1{\noindent\vbox{\chrule\hbox{\cvrule\kern3pt\vbox{
	\advance\hsize-7pt\vskip-\parskip\kern3pt\bf#1
	\hbox{\vrule height0pt depth\dp\strutbox width0pt}
	\kern3pt}\kern3pt\cvrule}\chrule}}




\def\today{\ifcase\month\or
 January\or February\or March\or April\or May\or June\or July\or
 August\or September\or October\or November\or December\fi
 \space\number\day, \number\year}

\parindent=20pt
\newskip\normalparskip	\normalparskip=.7\medskipamount
\parskip=\normalparskip	



\catcode`\|=\active \catcode`\<=\active \catcode`\>=\active 
\def|{\relax\ifmmode\delimiter"026A30C \else$\mathchar"026A$\fi}
\def<{\relax\ifmmode\mathchar"313C \else$\mathchar"313C$\fi}
\def>{\relax\ifmmode\mathchar"313E \else$\mathchar"313E$\fi}


%
%
%
%
%
%
%

\def\thetitle#1#2#3#4#5{
 \def\titlefont{\biggest} \font\footrm=cmr10 \font\footit=cmti10
  \twelverm
	{\hbox to\hsize{#4 \hfill YITP-SB-#3}}\par
	\vskip.8in minus.1in {\center\baselineskip=2.2\normalbaselineskip
 {\titlefont #1}\par}{\center\baselineskip=\normalbaselineskip
 \vskip.5in minus.2in #2
	\vskip1.4in minus1.2in {\twelvebf ABSTRACT}\par}
 \vskip.1in\par
 \narrower\par#5\par\unnarrower\vskip3.5in minus3.3in\eject}
\def\paper\par#1\par#2\par#3\par#4\par#5\par{
	\thetitle{#1}{#2}{#3}{#4}{#5}} 
\def\author#1#2{#1 \vskip.1in {\twelveit #2}\vskip.1in}
\def\YITP{C. N. Yang Institute for Theoretical Physics\\
	State University of New York, Stony Brook, NY 11794-3840}
\def\WS{Warren Siegel\footnote{$*$}{
	\pdflink{mailto:siegel@insti.physics.sunysb.edu}\\
	\pdfklink{http://insti.physics.sunysb.edu/\~{}siegel/plan.html}
	{http://insti.physics.sunysb.edu/\noexpand~siegel/plan.html}}}


\pageno=0

\paper

{\rgb{0.3 0 1}{Deriving projective hyperspace from harmonic}}

\author{Dharmesh Jain\footnote{${}^{\ddagger}$}{%
	\pdflink{mailto:Dharmesh.Jain@stonybrook.edu}} 
	and \WS}\YITP

08-7

March 20, 2009

We derive actions for projective N=2 superspace (``hyperspace") from those for harmonic hyperspace, including that for nonabelian Yang-Mills (a new result).  The method uses Wick rotation of the sphere from complex conjugate coordinates to real, null ones, which can be treated as independent.  The result can be considered ``holographic" in that the dimension of the internal (R-symmetry) space is reduced from 2 to 1, by solving equations of motion or gauge conditions for dependence on the other coordinate.  The auxiliary nature of the redundant dimension makes the hypergraph rules and evaluation almost identical.

\pageno=2

Ü1.  Introduction

There are two competing, but closely related, formalisms for effectively dealing with N=2 supermultiplets (``hypermultiplets" [1]) in N=2 superspace (``hyperspace"), in 4 dimensions (or simple supersymmetry in 6):  projective [2] and harmonic [3].  Projective hyperspace has the advantage of 1 less R-symmetry coordinate, which results in all the coordinates fitting neatly into a square matrix, whose hyperconformal transformations take the form of fractional linear transformations [4,5], hence the term ``projective".

Although the relations between various multiplets in the two formalisms has been frequently discussed, in this paper we will provide a direct derivation of multiplets, gauge transformations, and actions for the projective formalism from those of the harmonic.  The derivation is mostly straightforward:  The basic step is to start with the usual complex CP(1) coordinates $y$ and $Ðy$ of the (2-)sphere, which is the space of the SU(2)(/U(1)) R-symmetry, and treat them as independent, which can be accomplished by Wick rotation.  Due to the change in topology from compact to non-compact, the standard equations of motion and gauge conditions of the harmonic formalism, which involve only the (SU(2)- and gauge-)covariant $Ðy$ derivatives, no longer put the theory on shell.  We solve the equations of motion or gauge conditions for explicit $Ðy$ dependence of the hyperfields in terms of ``coefficients" that depend on $y$, and perform the $Ðy$ integral in the action.  (Instead of gauge fixing we can also define the projective gauge field in terms of the line integral of the harmonic one across the range of $Ðy$.)  Effectively, the theory has been reduced to its ``boundary" in $Ðy$, for both the hyperfields and their residual gauge invariance.  This is not the true boundary of the Wick-rotated theory, but symmetry under finite SU(2) R transformations is maintained on this one-dimensional space $y$.  (The exception is projective actions for nonrenormalizable theories that require integration over a specific contour in their definition, such as for the tensor multiplet, past which SU(2) can move singularities.  Such theories are SU(2) invariant under infinitesimal, but not finite transformations.  They also do not have SU(2) covariant forms in the harmonic formalism.)  This Wick rotation also accounts for the modified definition of charge conjugation used in such spaces [6].  The remaining coordinate can consistently be treated as real, even though the SU(2) transformations are complex, by treating them as being on the fields, rather than on the coordinates, since the fields that appear in the Laurent expansion in $y$ are complex (but may be subject to reality conditions based on charge conjugation consistent with SU(2)).

Previously [7], the equivalent was accomplished by replacing regular functions on the sphere with singular functions there in the harmonic formalism (or by taking a singular limit of regular functions), which allowed projective multiplets to be obtained after minor modifications, but altered the harmonic interpretation.  Here we do not modify the definition of the harmonic fields or action; the singularities of the projective fields in $y$ follow directly from the regular harmonic expansion of the harmonic fields.

We also give further analysis of hypergraphs in the 2 formulations, and evaluate the 1-scalar-hypermultiplet-loop divergence with an arbitrary number of external (nonabelian) vector-multiplet lines.  In particular, we give for the first time the complete projective action for nonabelian hyper Yang-Mills, which could be guessed from the similar harmonic action.

Ü2.  Coordinates and integrals

We begin with some conventions, and our definition for evaluation of the simple $Ðy$ integrals that convert harmonic hyperfields to projective ones.  We will ignore questions of representation with respect to the usual superspace coordinates until later, and focus mostly on just R-space.  We begin with a conventional parametrization of an element of SU(2) as
$$ g = \pmatrix{ e^{iÄ/2} & 0 \cr 0 & e^{-iÄ/2} \cr} {1\over å{1+yÐy}} 
	\pmatrix{ 1 & -Ðy \cr  y & 1 \cr} ­ \pmatrix{ Ðu \cr u \cr}$$

\noindent where the angle $Ä$ parametrizes the element of U(1) factored out to leave the projective complex conjugate coordinates $y$ and $Ðy$ of the sphere.  The currents $g^{-1}dg$ and $(dg)g^{-1}$ then define the dual SU(2) generators $G$ and covariant derivatives $d$, respectively, as usual, as
$$ \li{ G_0 & = y»_y - Ðy»_{Ðy} - i»_Ä \cr
	G_y & = »_y + Ðy^2 »_{Ðy} + iÐy »_Ä \cr
	G_{Ðy} & = y^2 »_y + »_{Ðy} - iy»_Ä \cr} $$
$$ \li{ d_0 & = - i»_Ä \cr
	d_y & = e^{iÄ} \left[ (1+yÐy) »_y - iÐy »_Ä \right] \cr
	d_{Ðy} & = e^{-iÄ} \left[ (1+yÐy) »_{Ðy} + iy»_Ä \right] \cr} $$

We then make the change of variables
$$ Ðy¼£¼t = {1\over 1+yÐy} $$
The convenience can be seen from the change of (Haar) measure for the coset (sphere):

$$ {1\over 2¹i}Ç{dyÊdÐy \over(1+yÐy)^2}¼£¼{1\over 2¹i}Ç{dy\over y}Ç_0^1 dt $$

\noindent (normalized so the integral of 1 is 1). 
At this point we are already effectively treating $y$ and $Ðy$ (now $t$) as Wick rotated coordinates, so they can be integrated independently.  (This corresponds to independent deformations of contours of integration of the 2 real coordinates of the sphere.)  The triviality of the measure for $t$ implies that covariant differential equations in that coordinate will also be.  The range of $t$ follows from the positivity of $yÐy$ on the sphere; we'll keep this restriction after Wick rotation to reproduce the usual projective hyperspace formalism.  (Although extending the range to the boundary of the Wick-rotated space at $t=¥$ should lead to the usual holography, we have not been able to derive a corresponding hyperspace formalism.)  The $y$ integral will then be interpreted as a (closed) contour integral. (Reality conditions will be discussed below.)

Another useful change of variables is
$$ e^{-iÄ}¼£¼e^{-i\Ä} = e^{-iÄ}t  $$
$$ Üâ»_\Ä = »_Ä $$
so $d_0$ is still integer or half-integer.  (A similar variable was used, with $y$ and $Ðy$, in [7].)  After switching from harmonic to projective hyperspace, this complex redefinition allows the complex gauge condition $\Ä=0$.  Another interpretation is to replace the R-sphere with a true CP(1): 2 complex coordinates with a complex scale invariance, allowing metrics that differ from the sphere by a Weyl scale (including flat R-space).  Then $\Ä=0$ is a choice of that complex scale.

So our final parametrization of the group element is
$$ g = \pmatrix{ e^{i\Ä/2} & 0 \cr 0 & e^{-i\Ä/2} \cr} 
	\pmatrix{ t & -(1-t)/y \cr  y & 1 \cr} $$

\noindent The symmetry generators and covariant derivatives are now
$$ \li{ G_0 & = y»_y - i»_\Ä \cr
	G_y & = »_y - {1\over y}(1-t)»_t \cr
	G_{Ðy} & = y^2 »_y - y ( t»_t + 2i»_\Ä ) \cr} $$
$$ \li{ d_0 & = - i»_\Ä \cr
	d_y & = e^{i\Ä}\left[ »_y - {1\over y}(1-t)( t»_t + 2i»_\Ä )\right] \cr
	d_{Ðy} & = - e^{-i\Ä} y »_t  \cr} $$

Determination of $Ðy$ dependence of hyperfields is simple, since the free field equations or gauge conditions we solve take the form $d_{Ðy}=0$ or $d_{Ðy}{}^2=0$, so the harmonic hyperfield consists of 1 or 2 projective ones by simple Taylor expansion in $t$.  Together with the determination of $\Ä$ dependence by the isotropy constraint, which determines the eigenvalue of $d_0$ for the harmonic hyperfield $ï$, we find
$$ d_0 ï = nï,â(d_{Ðy})^m ï = 0âÜâï = e^{in\Ä} Ý_{j=0}^{m-1} Æ_j(y) t^j $$
The analyticity properties of the projective hyperfields $Æ_j$ in $y$ then follow from the regularity of the original (off-shell) $ï$ on the sphere; we'll discuss each case individually below.  Since the field equations are no higher than second order in derivatives, the projective hyperfields can be associated with ``boundary values" (at $t$ = 0 or 1) of the harmonic ones.

The usual charge conjugation of the projective and harmonic formalisms (with respect to just SU(2); again we ignore the generalization to the full projective hyperspace [6]) is defined by the pseudoreality of the defining representation of SU(2), as given here by the group element (with respect to just the symmetry group):  Left-multiplication of $g*$ (where ``$Ê{}*Ê$" is ordinary complex conjugation) by an antisymmetric matrix gives back the same representation.  So
$$ Cu=ÐuâÜâ(Cy)* = -{1\over y},â(Ct)* = 1 - t,â(C e^{-i\Ä/2})* = y e^{-i\Ä/2} $$
Thus in projective hyperspace, which doesn't have $t$, $C$ switches a projective hyperfield associated with $t=0$ with one associated with $t=1$.  (I.e., in terms of initial and final ``time" $t$, it is time reversal.)  So from the above solution in terms of projective hyperfields $Æ_j$ of the field equations on a harmonic hyperfield $ï$ we have,
$$ (Cï)(z) = [ï(Cz)]*âÜâ(CÆ)(z) = y^{-2n}[Æ(Cz)]* $$
(We include all coordinates in $z$, so $C$ acts also on $x$ and $Ï$, which we haven't discussed.)  Hyperfields that have integer eigenvalue of $d_0$ are called ``real" if they are equal to their charge conjugates (whereas half-integer ones are pseudoreal representations of SU(2)).

Ü3.  Scalar hypermultiplet

Our general procedure for deriving projective actions from harmonic ones is to solve the field equation (for scalar multiplets) or the gauge condition (for the vector multiplet), both of which involve $t$-derivatives, and plug the solution back into the action.

For scalar multiplets the procedure is similar to the JWKB approximation in the path-integral formalism:  The ``classical" contribution is given by substituting the solutions of the equations of motion in terms of the boundary values (at ``initial" and ``final" times).  In our case, these boundary values of the harmonic hyperfields at $t=0$ and $1$ are the projective hyperfields.

There are two versions of the scalar hypermultiplet in harmonic hyperspace, but both reduce to the same one in projective hyperspace.  The one that's easier to treat is also the one that appears for the usual Faddeev-Popov ghosts:  Its free Lagrangian is
$$ L_1 = ü(d_{Ðy}¿)^2,âd_0 ¿ = 0,âC¿ = ¿ $$
As described in the previous section, the solution to its field equation is
$$ ¿ = ¿_0(y) + t¿_1(y)âÜâd_{Ðy}Ê¿ = -e^{-i\Ä}y¿_1(y) $$
In terms of the boundary values,
$$ ¿_i(y) = ¿|_{t=0} = ¿_0,â¿_f(y) = ¿|_{t=1} = ¿_0 + ¿_1 $$
$$ Üâ¿ = (1-t)¿_i + t¿_f $$
we find the reality condition
$$ ¿_f(y) = (¿_i)ÿ(-\f1y) $$

Regular functions on the sphere can be expanded in terms of spherical harmonics, or equivalently in terms of U(1)-invariant products of the SU(2) group element.  In the present case (integer isospin), these can be obtained from symmetrized products of those for isospin 1, namely
$$ {(y,Ê Ðy,Ê 1-yÐy)\over 1+yÐy} = \left( ty,Ê {1-t\over y},Ê 2t-1 \right) = 
	\cases{ ( 0,Ê {1\over y},Ê -1 ) & for $t=0$ \cr ( y,Ê 0,Ê 1 ) & for $t=1$ \cr} $$
So such harmonic fields will have only nonpositive powers of $y$ at $t=0$ ($¿_i$), and only nonnegative powers at $t=1$ ($¿_f$).  This is just the usual definition of the scalar multiplet $ç$ in projective hyperspace, regular at $y=0$, so we identify
$$ ç = ¿_f,âÑç ­ (ç)ÿ(-\f1y) = ¿_i $$

The projective Lagrangian is then the usual
$$ Ç_0^1 dt¼L_1 = -e^{-2i\Ä}y^2\left(Ç_0^1 dt\right)Ñçç
	= - e^{-2i\Ä}y^2 Ñçç $$
The fermionic coordinates cancel the $\Ä$ dependence.  The $1/y$ in the harmonic measure reduces the $y^2$ factor to $y$, a weight factor for charge conjugation [6], as the Lagrangian is a hyperconformal density in projective hyperspace.  We have dropped the $ç^2$ and $Ñç^2$ terms, which vanish after $y$ (and $Ï$) integration from lack of $1/y$ poles.

The other version of the free scalar hypermultiplet is described by the Lagrangian
$$ L_2 = Ðq d_{Ðy} q,âd_0 q = -ü q $$
(where $Ðq=Cq$).  The solution to its field equation is [7]
$$ q = e^{-i\Ä/2}q_0(y),âÐq = e^{-i\Ä/2}y(q_0)ÿ(-\f1y) ­ e^{-i\Ä/2}yÐq_0(y) $$

The Lagrangian would then seem to vanish, but we know from path integrals for fermions in quantum mechanics that a more careful, discretized-``time" analysis can lead to nonvanishing results, depending on the boundary conditions.  In particular, for a first-quantized Lagrangian of the form $ÐÆÀÆ$, time independence of $Æ$ and $ÐÆ$ by the equations of motion implies that the propagator gives just the inner product (i.e., the same result as $t_f=t_i$).  So, if the boundary conditions are chosen so that the initial wave function depends on $ÐÆ$ while the final depends on the canonical conjugate $Æ$, the ``classical" action found from the JWKB expansion is just $ÐÆ_i Æ_f$, whose exponentiation gives the ``plane wave" inner product.  Effectively, the result is the same as dropping the derivative, as for a ``boundary term" that might result on integration by parts.

In this case, this leads to the result
$$ Ç_0^1 dt¼L_2 = Ðq[-e^{-i\Ä}y]q = -e^{-2i\Ä}y^2 Ðq_0 q_0 $$
which is again the projective scalar hypermultiplet action, identifying
$$ ç=q_0 $$
The regularity of $ç$ at $y=0$ follows from associating $q_0$ with the original $q$ at $t=1$, and $Ðq_0$ with $t=0$.

(4D massive scalar hypermultiplets are found from 6D massless by dimensional reduction.)

Ü4.  Vector hypermultiplet

Unlike the scalar hypermultiplets, the reduction of the vector hypermultiplet follows from applying the gauge condition, rather than the field equation.  Solving the gauge condition is equivalent to (but more convenient than) working directly in terms of gauge-invariant variables.  The residual gauge invariance (in either method) is that of the projective formalism:  The gauge condition trivializes $Ðy$ dependence in both the gauge field and the gauge parameters.

Again from the above analysis, solving the usual gauge condition gives [7]
$$ d_{Ðy}A_{Ðy} = 0,¼d_0 A_{Ðy} = - A_{Ðy}âÜâA_{Ðy} = -ie^{-i\Ä}yV(y) $$
where we have defined $A_{Ðy,0}=yV$ by analogy with $d_{Ðy}$.  ($V$ is Hermitian with respect to $C$.) In the Abelian case, using the covariant current
$$ J^{Ðy} = dÐy¼e^{i\Ä}t^2 = dt¼e^{i\Ä}{1\over y} $$
(from $(dg)g^{-1}$), where $J^{Ðy}d_{Ðy} = dÐyÊ»_{Ðy} = dtÊ»_t$, to define the covariant line integral
$$ Abelian:âV ­ iÇ_0^1 J^{Ðy}A_{Ðy} = Ç_0^1 dt¼V = V $$
we see the gauge-independent definition of $V$ is consistent with the above gauge condition.  For the nonabelian case, we instead define the (complexified) group element
$$ e^V ­ \P \left[ exp \left( iÇ_0^1 J^{Ðy}A_{Ðy} \right) \right] $$
again consistent with the above gauge.  ($C$ gives an extra sign change from switching $tª1-t$, so hermitian conjugation with $C$ gives 2 canceling path reversals.)

The regularity of $A_{Ðy}$ (in arbitrary gauges) tells us it has the above type of singularities in $y$ at $t=0$ or $1$.  Thus, $V$ must have singularities at both $y=0$ and $¥$, as in the usual projective formalism.  Furthermore, examining the abelian gauge transformation applied to the gauge-independent definition of $V$ as a line integral
$$ Abelian:â¶A_{Ðy} = - d_{Ðy}K,¼d_0 K = 0âÜâ-i¶V = K|_{t=0}^1 $$
and using the correspondence between the scalar multiplet $¿$ and gauge parameter $K$ in harmonic hyperspace on the one hand, and the scalar multiplet $ç$ and gauge parameter $ñ$ in projective hyperspace on the other hand (except for different conformal weights), we recognize the usual Abelian projective gauge transformation
$$ Abelian:â¶V = i(ñ - Ðñ) ;âñ = K|_{t=1},¼Ðñ = K|_{t=0} $$
Because of the path ordering in the gauge-independent definition, this can be seen to generalize directly to the nonabelian case as
$$ e^{V'} = e^{-iÐñ}e^V e^{iñ} $$

Ü5.  Vector hypermultiplet coupling

Before looking at the action, we examine the coupling to matter.  In the above gauge, even in the nonabelian case, the $Ðy$ covariant derivative can be written as
$$ á_{Ðy} = d_{Ðy} + iA_{Ðy} = e^{tV}d_{Ðy}e^{-tV} $$
This modifies the solution to the matter field equations: e.g.,
$$ ¿ = e^{tV} ( ¿_0 + t¿_1 )âÜâ¿_i =¿_0,¼¿_f = e^V ( ¿_0 + ¿_1 ) $$
$$ Üâ¿ = e^{tV} (1-t) ¿_i + e^{-(1-t)V} t ¿_f $$
$$ Üâd_{Ðy}Ê¿ = - e^{-i\Ä}e^{tV}y¿_1 = e^{-i\Ä}y( e^{tV}¿_i - e^{-(1-t)V}¿_f ) $$
Since $¿$ must be a real representation of the Yang-Mills group, the group generators are antisymmetric, so
$$ L_1 = ü(d_{Ðy}¿)^T d_{Ðy}¿ = - e^{-2i\Ä}y^2 ¿_f e^V ¿_i $$
again after dropping non-cross terms, whose $V$ dependence cancels, and so vanish after integration as before.  The result is the usual modification by $e^V$, which restores gauge invariance.  If we write $e^V$ as a gauge-covariant path-ordered exponential of the integral of $A_{Ðy}$, we recognize this modification as gauge-covariant point splitting in $t$.  Similarly, for the other multiplet we have [7]
$$ q = e^{-i\Ä/2}e^{tV}q_0,âÐq = e^{-i\Ä/2}yÐq_0 e^{(1-t)V} $$
yielding the same result.

The nonabelian gauge transformation of $V$ can be derived from the above expression for that of $e^V$.  An alternate method is to solve for the residual gauge invariance in the above gauge.  This is equivalent to solving the equations of motion for the Faddeev-Popov ghosts.  The equation to solve is
$$ 0 = ¶(d_{Ðy}A_{Ðy}) = -d_{Ðy}[á_{Ðy},K] $$
Plugging in the above expression for $A_{Ðy}$ in this gauge yields 
$$ »_t e^{tV} »_t e^{-tV} K = 0 $$
where we now write $K$ as a column vector (so $V$ is in the adjoint representation) for convenience.  The solution, in notation analogous to that for $¿$ above, is
$$ K = e^{tV}K_0 + {1\over V} ( e^{tV} - 1 ) K_1
	= {1\over e^V - 1} \left[ (e^V - e^{tV}) K_i + (e^{tV} - 1) K_f \right] $$
(Upon Taylor expansion, there are no inverse powers of $V$.)  The transformation law is then
$$ ¶V = -iüV \left[ ( Ðñ + ñ ) + coth (üV) ( Ðñ - ñ ) \right] $$
in analogy to the N=1 result.

In an arbitrary gauge, we have
$$ á_{Ðy} = \P \left[ exp \left( iÇ_0^t J^{Ðy}A_{Ðy} \right) \right] d_{Ðy}¼
	\P \left[ exp \left( iÇ_t^0 J^{Ðy}A_{Ðy} \right) \right] $$
if we assume the boundary condition
$$ A_{Ðy}|_{t=0} = 0 $$
(This might also be an asymptotic gauge condition, but it seems reasonable as a boundary condition since $d_{Ðy}$ has a factor of $1/t$ multiplying $»_{Ðy}$.)  This uses the explicit gauge transformation for going to the gauge $A_{Ðy}=0$.  Repeating the above manipulations then produces the same results but in terms of the gauge-covariant definition of $V$ given above.  This construction is reminiscent of the construction for N=1, where $e^V=e^¯ e^{Я}$, with $Я$ corresponding to the $Ç_0^t$ piece and $¯$ to the $Ç_t^1$.  This allows transformations to different gauge representations where the covariant derivatives transform with only one of $K­K(t)$, $ñ­K(1)$, or $Ðñ­K(0)$.

Ü6. Fermion representations

Representations with respect to spinor derivatives differ slightly in the 2 formalisms because of the (non)appearance of $Ðy$.  Just as the covariant R-derivatives of the harmonic formalism are invariant under the global SU(2) (commute with the generators), the usual covariant spinor derivatives need to be multiplied by the group element $g$ to replace their SU(2) transformations with those of the isotropy U(1):
$$ \pmatrix{ d_Ï \cr d_\Ï \cr } = g\pmatrix{ d_{(1)} \cr d_{(2)} \cr }âÜâ
	d_\Ï = e^{-iÄ/2}åt( d_{(2)} + yd_{(1)} ),â
	d_Ï = e^{iÄ/2}åt( d_{(1)} - Ðy d_{(2)} ) $$
where $d_\Ï$ vanishes on projective hyperfields.  Here we use six-dimensional SU*(4) matrix notation for spinors (and vectors):  In the ``real" representation, $d_{(1)}$ and $d_{(2)}$ are hermitian conjugates of each other up to an antisymmetric 4$ð$4 matrix; they form the usual pseudoreal isospinor representation of the global SU(2).  Their anticommutation relations are
$$ Ód_{(1)},d_{(2)}Õ = - Ód_{(2)},d_{(1)}Õ = -i»_x,ââÓd_{(1)},d_{(1)}Õ = Ód_{(2)},d_{(2)}Õ = 0 $$
where the sign is due to the antisymmetry of the 4$ð$4 matrix $»_x$ (6 coordinates for D=6, but easily reduced to D=4).  

In terms of our redefined SU(2) coordinates,
$$ d_\Ï = e^{-i\Ä/2}( d_{(2)} + yd_{(1)} ),â
	d_Ï = e^{i\Ä/2}\left[ td_{(1)} - (1-t){1\over y}d_{(2)} \right] $$
Clearly $d_Ï$ needs to be redefined for the projective formalism:  Fixing any value of $t$ will preserve the spinor-derivative anticommutation relations; $t=1$ is the choice that relates directly to the usual projective formalism, as well as giving the simplest $y$ dependence.  Similar remarks apply to $d_y$.

The real representation is the least useful one for the projective formalism.  The representations that are more useful are related to ``twisted-chiral" representations in the original superspace, obtained by supercoordinate transformations $x£xàüi\ÏÏ$:
$$ d_{(1)} = »_Ï +i\Ï »_x,âd_{(2)} = »_\Ïâorâd_{(1)} = »_Ï,âd_{(2)} = »_\Ï -iÏ»_x $$
The former leads to the ``analytic" representation in the harmonic formalism after a further redefinition involving the R-coordinates $Ï£Ïà\Ï y$.  After manipulations like the above, similar (but not identical) representations can be obtained for projective hyperspace.

However, the desired representations can be both obtained and explained more directly in projective hyperspace:  We first note that the (4D) hyperconformal group can be represented directly on the projective coordinates via fractional linear transformations (as for other projective spaces, such as SU(2) on CP(1)).  Under this representation of the hyperconformal group, simple translations of the coordinates yield the usual $x$ translations, half the hypersymmetries, and some of the R-symmetry.  We call this the ``projective representation".  But there is another representation where it is the corresponding covariant derivatives that are just partial derivatives, instead of the generators of this subgroup of the hyperconformal group.  The existence of this other representation is clear if we consider the hyperspace coordinates in terms of hyperconformal group elements.  At first we ignore the isotropy group, which is generated by a subset of the covariant derivatives.  Then there is a symmetry between hyperconformal generators and covariant derivatives as they are generated by left and right action on the group element.  These representations can easily be switched by the coordinate transformation that replaces the group element by its inverse:
$$ g' = g_L g g_RâÜâ(g^{-1})' = g_R^{-1}g^{-1}g_L^{-1} $$
$$ g £ g^{-1}âÜâg_L £ g_R^{-1},âg_R £ g_L^{-1} $$
In practice, it's more convenient to replace this transformation with one that can be obtained continuously from the identity, by in addition performing a sign change for all the coordinates.  These 2 transformations would cancel for exponential parametrization of the group element.  But for the more standard parametrization as a ÓproductÕ of exponentials, this combination just reverses the ordering of the exponential factors.  In this case, it is equivalent to a hyperconformal transformation on the projective coordinates (and not $\Ï$) with $\Ï$ acting as the parameter, of the form described above.  

The resulting ``reflective" representation is essentially one of the twisted-chiral representations described above (with $t£1$).  The projective representation is like the other one, but requires in addition a $y$-dependent hypercoordinate transformation.
The net result for the covariant derivatives $d$ and corresponding symmetry generators $G$ of the 2 representations is

$$ \def\normalbaselines{\advance\baselineskip1\jot}
	\matrix{ & projective & reflective \cr
	d_x & »_x & »_x \cr
	d_Ï & »_Ï +i\Ï »_x & »_Ï \cr
	d_y & »_y - \Ï »_Ï -i ü \Ï\Ï »_x & »_y \cr
	d_\Ï & »_\Ï & »_\Ï + y»_Ï -iÏ»_x  \cr
	G_x & »_x & »_x \cr
	G_Ï & »_Ï & »_Ï -i\Ï »_x \cr
	G_y & »_y & »_y + \Ï »_Ï -i ü \Ï\Ï »_x \cr
	G_\Ï & »_\Ï - y»_Ï +iÏ»_x & »_\Ï  \cr} $$

The advantages of the projective representation are that there, projective hyperfields depend on just the projective coordinates, hyperconformal transformations are simpler, and scattering amplitudes are simpler because their hyperspace form (as derived, e.g., from hypertwistors) contains explicit hypersymmetry conservation $¶$-functions $¶(ÝG_Ï)$ for $G_Ï=»_Ï$.  The advantage of the reflective representation is that the $y$-nonlocal action for hyper Yang-Mills (see below) can be written simply.  (The same is true for gauge-covariant derivatives, written in a similar form.)  The corresponding expressions in the projective representation are more complicated, because the $y$-dependent transformation from a real (or reflective) representation to the projective one (which isn't needed from real to reflective) is different at each $y$.  This is related to the fact that such actions have explicit $\Ï$-dependence.  However, it is possible to perform the $\Ï$ integration; the result contains derivatives in a form that is not manifestly covariant.  (By analogy, consider an N=1 action of the form $Çd^4 ÏÊL(Ä,d_Œ Ä)$ depending only on the chiral $Ä$ and not antichiral $ÐÄ$.)

Ü7. Hypergraphs

A few N=2 supergraphs have been evaluated in both approaches.  The rules and tricks were similar, due to the fact that the harmonic formalism [8] differs from the projective one [9] only by the appearance of additional auxiliary multiplets, coming from extra $Ðy$ (or $t$) dependence.  We summarize these rules here in our notation.  Those that are (almost) the same are (in real/reflective representations, or those that differ by only $y$-independent coordinate transformations):
$$ \li{ \hbox{scalar multiplet propagator:}&â
	{d_{1\Ï}^4 d_{2\Ï}^4\over y_{12}^3}{¶^8(Ï_{12})\over p^2} \cr
\hbox{vector multiplet propagator:}&âd_\Ï^4 ¶(y_{12}){¶^8(Ï_{12})\over p^2}â
	\hbox{(Fermi-Feynman gauge)} \cr
\hbox{scalar multiplet vertex:}&âÇd^4 ÏÊdy,â\hbox{but use} Çd^4 ÏÊd_\Ï^4 = Çd^8 Ï \cr
\hbox{vector multiplet (only) vertex:}&âÇd^8 ÏÊdy_1...dy_n{1\over y_{12}y_{23}...y_{n1}} \cr} $$

\noindent where $Ï_{12}­Ï_1-Ï_2$, etc.  (The rules above are for the $q$ scalar multiplet in the harmonic formalism, which is simpler, and more similar to the projective case.  The result for the vertex for self-interacting vector multiplets for the projective formalism is given by analogy to the harmonic, and will be derived below.)  There are also the identities common to both:
$$ d_{2\Ï}d_{1\Ï}^4 = y_{21}d_{1Ï}d_{1\Ï}^4âÜâ
	¶^8(Ï_{12}) d_{2\Ï}^4 d_{1\Ï}^4 ¶^8(Ï_{12}) = y_{12}^4 ¶^8(Ï_{12}) $$
The former is used when integrating a spinor derivative by parts from one propagator across a vertex to an adjacent propagator; alternatively, the latter can be used when only 4 such derivatives are moved in the last step of $Ï$ integration.

The differences in the above expressions in the two formalisms are the number of R coordinates and the $i·$ prescription:
$$ \def\normalbaselines{\advance\baselineskip1\jot}
	\matrix{ & \hbox{harmonic} & \hbox{projective} \cr
	``Çdy" & ÇdyÊdÐy/2¹i(1+yÐy)^2 & Èdy/2¹i \cr
	``¶(y_{12})" & 2¹i(1+yÐy)^2 ¶(y_{12})¶(Ðy_{12}) & 2¹i¶(y_{12}) \cr
	``1/y_{12}" & 1/(y_{12}+·/Ðy_{12}) & 1/[y_{12}-·(y_1+y_2)] \cr} $$

\noindent In manipulations involving integrating ``$1/y$" to make results more R-local, in the harmonic formalism one needs various identities that generate $Ðy$ derivatives to apply 
$$ »_{Ðy}Ê{1\over y+·/Ðy} ¾ ¶^2(y) $$
which is easy to integrate.  On the other hand, in the projective formalism one just immediately evaluates standard contour integrals.  There is also an ordering for the projective formalism:  $1/[y_{12}-·(y_1+y_2)]$, for $y_1$ and $y_2$ on the same contour, is for $Òç(1)Ðç(2)Ô$.  This means that effectively one integrates with the $y_2$ contour enclosing $y_1$ (and 0), or the $y_1$ contour inside $y_2$ (and $¥$).  For example, for contours counterclockwise around the origin, we have

$$ {1\over y_{12}-·(y_1+y_2)} + {1\over y_{21}-·(y_1+y_2)} = -2¹i¶(y_{12}) $$

Another source of differences is the relation of the spinor derivatives in the 2 approaches:  We have seen that the projective ones follow from the harmonic ones effectively by setting $t=1$.  (We also gauge $\Ä=0$ in both cases.)  So for the harmonic relations
$$ Ód_{1\Ï},d_{2\Ï}Õ = -i u_1Éu_2 »_x,âÓd_{1Ï},d_{2\Ï}Õ = -i Ðu_1Éu_2 »_x,â
	Ód_{1Ï},d_{2Ï}Õ = -i Ðu_1ÉÐu_2 »_x $$
we have in general
$$ u_1Éu_2 = y_{12},âÐu_1Éu_2 = t_1 +(1-t_1){y_2\over y_1},â
	Ðu_1ÉÐu_2 = t_2 (1-t_1){1\over y_1} -  t_1 (1-t_2){1\over y_2} $$
but only for the projective case do the latter 2 simplify:
$$ u_1Éu_2 = y_{12},âÐu_1Éu_2 = 1,âÐu_1ÉÐu_2 = 0â\hbox{(projective)} $$
$$ ÜâÓd_{1\Ï},d_{2\Ï}Õ = -i y_{12} »_x,âÓd_{1Ï},d_{2\Ï}Õ = -i »_x,âÓd_{1Ï},d_{2Ï}Õ = 0 $$
Moving spinor derivatives from propagators around loops requires evaluating expressions of the form
$$ d_{i\Ï}...d_{j\Ï}d_{1\Ï}^4 $$
which results in repeated use of the above anticommutators, so the harmonic formalism also has these $t$-dependent factors to deal with.  (The example above that gave the same result in the 2 approaches needed only $u_1Éu_2$.)  However, one should be able in general to use $d_{(1)}$ in place of $d_Ï$ in the harmonic approach to mimic the projective and get the same simplifications, since only $d_\Ï$ appears in the Feynman rules.

There is also some Legendre transformation involved in the ``duality", which accounts for the minor differences in the action, such as coupling to $iA_{Ðy}$ vs.¼$e^V-1$ (subtracting out the ``1" for the kinetic term).  Also, the rules for the $¿$ multiplet (and the ghosts) are a little more complicated than the $q$ multiplet for the harmonic formalism.  (For the most part, the extra $d_{Ðy}$ in the vertex converts the $¿$ propagator into a $q$ propagator.)

The bottom line is that although the final results in the 2 approaches are almost the same (to the same extent as the Feynman rules), the harmonic formalism requires some extra algebra (for R-space).

Ü8. Vector hypermultiplet action

One way to derive the action for the vector multiplet is from looking at the divergent part of a scalar multiplet loop in a vector background [8].  The calculation is almost the same in the two formalisms:  To keep the most divergent part, keep all spinor derivatives inside the loop when integrating them by parts, and keep the $»_x$ terms (vs.¼$yd_Ï^2$ terms) generated by pushing $d_\Ï$'s past $d_Ï$'s. Thus almost every $d_\Ï^4$ integrated by parts produces a $y^2 p^2$.    The result after performing all $Ï$ integration (except the usual final one) is that every $1/y^3$ is replaced by a $1/y$, while only 2 $1/p^2$'s remain (associated with the 2 $d_\Ï^4$'s killing the next-to-last $¶^8(Ï)$, as in the above identity), yielding the logarithmic divergence.

The main difference we have already seen:  While in the harmonic case $q$ couples to $iA_{Ðy}$, in the projective case $ç$ couples to $e^V-1$.  Doing the 1-loop calculation as just described, or just drawing this analogy to the harmonic case, the projective action is then

$$ S_{YM} ¾ tr Çd^4 xÊd^8 Ï Ý_{n=2}^¥ {1\over n} {dy_1\over 2¹i} ò {dy_n\over 2¹i}
	{ (e^{V(1)}-1) ò (e^{V(n)}-1)\over y_{12} ò y_{n1} } $$

\noindent There is also a ``dual" version, coming from reverse ordering of the loop, corresponding to starting with the action as $çe^{-V}Ñç$ rather than $Ñçe^V ç$.  The result is to everywhere change the signs on $V$ and $y$.  For such real representations $V^T=-V$, so transposing reproduces the above form.

The check of gauge invariance is similar to the harmonic case, but again no derivatives $d_{Ðy}$ are involved.  We start with
$$ ¶(e^V) = -iÐñ e^V + e^V iñâÜâ¶(e^V -1) = ( -iÐñ +iñ ) + [ -iÐñ (e^V -1) + (e^V -1) iñ ] $$
Then, as in the harmonic case, the inhomogeneous contribution to the ``$n$-point" (in $y$) contribution to the action will cancel the linear contribution to the ($n-1$)-point.  The exception is the inhomogeneous contribution to the 2-point, which vanishes by itself (after $Ï$ integration).  To see this cancelation, note that the projective $1/y$ acts as a 1D St¬uckelberg-Feynman propagator, propagating only ``positive-energy" modes (nonnegative powers of $y$) in one direction and ``negative-energy" modes (nonpositive powers) in the other (because in the above derivation it came from an $çÑç$ propagator).  The effect is that integration over the $y$ of an inhomogeneous contribution will result in attaching only $ñ$ to the right of the $e^V-1$ factor on its immediate left, and $Ðñ$ to the left of the $e^V -1$ on its right.  (However, one should not try to define each contour enclosing the previous simultaneously, implying a Penrose staircase.  Keeping all contours the same is consistent with the $i·$ prescription.)

Various alternative derivations of the action are possible:  One way would be to start with the harmonic action, choose the gauge as above, and explicitly integrate out the $t$ dependence, including that in the $i·$ prescription for the ``propagators".  (Note that the $y$ in each $iA_{Ðy}=yV$ cancels with that in $ÇdyÊdÐy/(1+yÐy)^2=ÇdtÊdy/y$.)  This is trivial for the kinetic term, because the $d^4\Ï$ integration can easily be performed:  Using $d_{1\Ï}^4$, it acts only on $V(2)$, generating terms with at least 2 factors of $y_{12}$, canceling the poles in $y$, and thus the $·$'s.  Since the $t$ integration is then trivial, and the result has the same form for the harmonic and projective formalisms, the equivalence is obvious.  (Similar remarks also apply for the finite part of the 1-scalar-hypermultiplet-loop correction to the vector-hypermultiplet propagator.)

Ü9. Background fields

It should be straightforward to develop a background-field formalism for the projective formalism, similar to the harmonic one [10].  An essential ingredient is the projective d'Alembertian $ßõ$,
$$ á_\Ï^4 á_y^2 á_\Ï^4 = üßõ á_\Ï^4 = á_\Ï^4 üßõ $$
$$ ßõ = õ +W^Œ á_Œ + \D_{Ðy} á_y +\D_0 $$
again in 6D notation, where $õ$ is the square of the background-covariantized $»_x$, $á_Œ$ is the covariantized $d_Ï$, and $á_y$ is the covariantized $d_y$.  Explicitly appearing field strengths are $W^Œ$ (physical spinors at $Ï=0$), and $\D_{Ðy}$ and $\D_0$ (auxiliary scalars at $Ï=0$).  On reduction to D=4, $õ$ also contributes a $ÓW,ÑWÕ$ term from the components of the gauge vector in the 2 extra directions, which become the chiral and antichiral scalar field strengths.  This d'Alembertian is the analog of the N=1 $á_\Ï^2 á_Ï^2 á_\Ï^2=üßõá_{\Ï}^2$ (for $á_Ï=á_Œ$ and $á_\Ï=Ñá_{ÀŒ}$).  $á_y^2=0$ is the (local) superconformal field equation that generates all others via commutators with $á_\Ï$.  (Similar remarks apply for the N=1 analog.)  Since its evaluation requires no $á_{Ðy}$, it can be used in the projective formalism in the same way as in the harmonic.  Furthermore, since it keeps superfields in projective (or harmonic analytic) hyperspace, all the $\Ï$-dependent pieces in the individual terms can be dropped, since they cancel.

Of particular interest is the fact that the vector multiplet propagator, in a background, can be expressed as the inverse of this operator.  Thus not only is the propagator local in $y$, as in ordinary (Feynman) gauges, but so are its 1-loop interactions.  As a result, all 1-loop contributions to the effective action from the vector multiplet (less ghosts) in a vector-multiplet background can be calculated directly in projective (or analytic harmonic) hyperspace, in the projective representation, without the appearance of $\Ï$.  In particular, in the N=2 formulation of N=4 Yang-Mills, the N=2 scalar multiplet cancels the (Faddeev-Popov and Nielsen-Kallosh) ghosts (in analogy to the N=1 case), so this contribution gives the complete result for the external vector multiplet.  (There is also a similar $ßõ$ contribution from the Nielsen-Kallosh ghost, which cancels all but a finite number of $y$ degrees of freedom, making the $y$ part of the loop trace finite.  In projective language, this is a cancelation of the ``arctic" pieces of a ``tropical" hyperfield.)

Thus, as for the N=1 formulation, not only do the contributions to the effective action from <4-point immediately vanish for N=4, but the expression for the 4-point is obvious:  As for N=1, 4 spinor derivatives are required inside the loop, and the final integral is over 4 $Ï$'s, so the result is proportional to the box integral of bosonic $\Ä^3$ theory times
$$ Çd^4 ϼW^Œ W^º W^© W^¶ ·_{Œº©¶} $$
Except for $y$-dependence, this is essentially the same as the N=1 result, where in the latter case $d^4 Ï$ is the integral over the full superspace, and $W^Œ$ is the Dirac 4-spinor combination of the chiral and antichiral Weyl spinor field strengths.  The above result also holds for the 6D theory (with $d^6 p$ for the momentum-space integral).  Although not manifestly projective, it can be evaluated at $\Ï=0$, as explained above.  In D=4 it can be re-expressed as an integral over the full superspace of the usual chiral and antichiral field strengths (as for N=1), which are scalars for N=2; but the above result may be more useful, as it applies also to D=6, and can be expressed and derived directly in projective hyperspace.  Another advantage of projective (or analytic) hyperspace is translation invariance in $Ï$, so Fourier transformation in the anticommuting coordinates is actually more convenient:  All the $¶(Ï)$'s can be avoided, and there is only 1 integral over loop fermions.

Note the similarity of this result to the case of N=4 superspace [11]:  There N=4 projective superspace (or maybe some harmonic analog) is the only convenient way to express this result (with a scalar field strength).  Also, in both cases the result is $y$-independent, so $y$ integration is redundant.  This suggests the possibility that in the N=4 case, where there is only the vector multiplet, all the supergraph rules might be formulated most conveniently in projective superspace.

Ü10. Conclusions

Our explicit relation between the projective and harmonic formalisms shows that in the appropriate notation the two are almost the same, sharing similar (dis)advan\-tages.  The only significant difference is the extra R coordinate of harmonic hyperspace, which appears in so simple a way as to have little effect. 

The one-loop form of the classical Yang-Mills action suggests including two new non-analytic hyperfields whose functional integration would generate it.  This action might be a Chern-Simons action that has been partially gauge fixed.

There is an N=3 harmonic formulation of N=4 Yang-Mills [12], but no amplitudes have been calculated with it.  It is possible that the corresponding projective formulation is already N=4:  The number of $Ï$'s (and $x$'s) is already the same, and the combination of field equation (since there are an infinite number of auxiliary fields) with gauge condition might reduce the N=3 harmonic's 6 R-coordinates to the N=4 projective's 4.  The N=3 action is curiously simpler than the N=2; this also suggests the existence of a simpler N=2 action.

ÜAcknowledgments

This work is supported in part by National Science Foundation Grant No. PHY-0653342.

\refs

£1 
  P. Fayet,
  ÓNucl.\ Phys.\  BÕ {\bf 113} (1976) 135.

£2 
  A. Karlhede, U. Lindstr¬om and M. Ro×cek,
  ÓPhys.\ Lett.\  B Õ{\bf 147} (1984) 297;
  \\
  U. Lindstr¬om and M. Ro×cek,
  ÓCommun.\ Math.\ Phys.\  Õ{\bf 115} (1988) 21,
  {\bf 128} (1990) 191.

£3 
  A. Galperin, E. Ivanov, S. Kalitsyn, V. Ogievetsky and E. Sokatchev,
  ÓClass.\ Quant.\ Grav.\  Õ{\bf 1} (1984) 469;
  \\
  A. Galperin, E. Ivanov, V. Ogievetsky and E. Sokatchev,
  ÓJETP Lett.\  Õ{\bf 40} (1984) 912
  [ÓPisma Zh.\ Eksp.\ Teor.\ Fiz.\  Õ{\bf 40} (1984) 155];
  \\
  B. M. Zupnik,
  ÓTheor.\ Math.\ Phys.\  Õ{\bf 69} (1986) 1101
  [ÓTeor.\ Mat.\ Fiz.\  Õ{\bf 69} (1986) 207];\\
A.S. Galperin, E.A. Ivanov, V.I. Ogievetsky, and E.S. Sokatchev, ÓHarmonic superspaceÕ
(Cambridge Univ. Press, 2001).

£4 
  W. Siegel,
  ÓPhys.\ Rev.\  D Õ{\bf 52} (1995) 1042
  \xxxlink{hep-th/9412011}.

£5 
  G.G. Hartwell and P.S. Howe,
  ÓInt.\ J.\ Mod.\ Phys.\  A Õ{\bf 10} (1995) 3901\\
  \xxxlink{hep-th/9412147},
  ÓClass.\ Quant.\ Grav.\  Õ{\bf 12} (1995) 1823;
  \\
  P. Heslop and P.S. Howe,
  ÓClass.\ Quant.\ Grav.\  Õ{\bf 17} (2000) 3743
  \xxxlink{hep-th/0005135};
  \\
  P.J. Heslop,
  ÓClass.\ Quant.\ Grav.\  Õ{\bf 19} (2002) 303
  \xxxlink{hep-th/0108235}.

£6 
M. Hatsuda and W. Siegel,
  ÓPhys.\ Rev.\  D Õ{\bf 77} (2008) 065017
  \xxxlink{0709.4605} [hep-th].

£7 
  S.M. Kuzenko,
  ÓInt.\ J.\ Mod.\ Phys.\  AÕ {\bf 14} (1999) 1737
  \xxxlink{hep-th/9806147}.

£8 
  A. Galperin, E. Ivanov, V. Ogievetsky and E. Sokatchev,
  ÓClass.\ Quant.\ Grav.\  Õ{\bf 2} (1985) 601,
  617.

£9 
  F. Gonzalez-Rey, M. Ro×cek, S. Wiles, U. Lindstr¬om and R. von Unge,
  ÓNucl.\ Phys.\  B Õ{\bf 516} (1998) 426
  \xxxlink{hep-th/9710250};
  \\
  F. Gonzalez-Rey and R. von Unge,
  ÓNucl.\ Phys.\  B Õ{\bf 516} (1998) 449\\
  \xxxlink{hep-th/9711135};
  \\
  F. Gonzalez-Rey,
  \xxxlink{hep-th/9712128};
  \\
  F. Gonzalez-Rey and M. Ro×cek,
  ÓPhys.\ Lett.\  B Õ{\bf 434} (1998) 303
  \xxxlink{hep-th/9804010}.

£10 
  I.L. Buchbinder, E.I. Buchbinder, S.M. Kuzenko and B.A. Ovrut,
  ÓPhys.\ Lett.\  B Õ{\bf 417} (1998) 61
  \xxxlink{hep-th/9704214};
  \\
  I.L. Buchbinder, S.M. Kuzenko and B.A. Ovrut,
  ÓPhys.\ Lett.\  B Õ{\bf 433} (1998) 335\\
  \xxxlink{hep-th/9710142};\\
  S.M. Kuzenko and I.N. McArthur,
  ÓPhys.\ Lett.\  BÕ {\bf 506} (2001) 140\\
  \xxxlink{hep-th/0101127};\\
  I.L. Buchbinder, E.A. Ivanov and A.Y. Petrov,
  ÓNucl.\ Phys.\  BÕ {\bf 653} (2003) 64
  \xxxlink{hep-th/0210241}.

£11 
  R. Kallosh,
  \xxxlink{0711.2108} [hep-th];\\
  M. Hatsuda, Y.-t. Huang and W. Siegel,
  \xxxlink{0812.4569} [hep-th].

£12 
  A. Galperin, E. Ivanov, S. Kalitsyn, V. Ogievetsky and E. Sokatchev,
  ÓPhys.\ Lett.\  B Õ{\bf 151} (1985) 215;
  \\
  F. Delduc and J. McCabe,
  ÓClass.\ Quant.\ Grav.\  Õ{\bf 6} (1989) 233.

\bye